\documentclass[11pt]{article} % 11pt font = 12pt Word font
\usepackage{pdfpages}
\usepackage{times}

\usepackage{geometry}
\usepackage{fullpage}

\begin{document}

\title{\bf
Experiments in Sustainable Software Practices for Future Architectures
}
\author{Charles R. Ferenbaugh
\thanks{
Mailing address:
Los Alamos National Laboratory, Mail Stop B295, Los Alamos, NM 87544, USA.
Email: {\tt cferenba@lanl.gov}.
\newline\indent
Los Alamos National Laboratory, an affirmative action/equal opportunity employer, is operated by Los Alamos National Security, LLC, for the National Nuclear Security Administration of the U.S. Department of Energy under contract DE-AC52-06NA25396.
\newline\indent
The author gratefully acknowledges the support of the NNSA Advanced
Simulation and Computing (ASC) Program, SWIFT project, for this work.
Thanks to Tim Kelley and Bryan Lally (project leads for Exa11 and Exa12, respectively) for helpful conversations.
This document is released internally at LANL as LA-UR-13-26936.
}
\\
HPC-1, Scientific Software Engineering \\
Los Alamos National Laboratory \\
}
\date{}
\maketitle
\thispagestyle{empty}

Since its beginnings during World War II, the nuclear weapons program
at Los Alamos National Laboratory has relied heavily on scientific
computation.
Large codes have been written over many years, and significant efforts
have been made to develop and implement cutting-edge physics methods.
However, relatively little attention has been given to
the computer science and software engineering aspects of the codes.
While some modern
software development practices have been adopted by the code teams
(version control, regression testing, automated builds and tests),
much more remains to be done.

In recent years, the code projects have given increased attention
to modern software development practices, due to a combination of factors.
Scientists who started out writing
small, temporary research tools for use by a few people are now working
on large, long-lived production codes with teams of developers
and dozens of users; this leads to an increased need for their code to be
maintainable and extensible.
Computational results that in the past were mainly used to inform
experiment design and analyze results are now
considered as deliverables in their own right; this places the code
that produced them under much greater scrutiny.
Projects that once could count on being well-funded now find their
budgets shrinking; this drives managers to try and use their
software development funds more efficiently.
All of these factors are leading LANL and many other scientific
computing centers to try to improve software development practices.

However, another driver has also arisen
that is perhaps more specific to LANL:  the need to deploy
our codes on future architectures such as many-core, GPU, 
Intel MIC, and so on.
The Roadrunner cluster was deployed at LANL in 2008.  It was the first
petaflop cluster, and also the first large-scale hybrid architecture,
containing both traditional x86\_64 CPUs and new IBM Cell processors
as accelerators.  Much effort was spent figuring out how to rewrite
our existing algorithms to perform well on this system.  Similar
efforts are ongoing to prepare our codes for other current and
future architectures.
During this process, we have found a number of areas in which
sustainable software practices can provide significant advantages.

\section*{Advantages of Sustainable Practices for Future Architectures}

In the area of code development, we found that many of our traditional
practices were not well suited to new architectures.  Our usual
coding style, best summarized as ``just get the physics working,''
led to code that was difficult to understand and modify.  In particular,
there was often heavy coupling between the physics algorithms themselves
and the infrastructure needed for performance (memory management,
database access, MPI calls, etc.); this made it hard to change
the infrastructure to support new architectures while
leaving the physics intact.  Also, our data structures were often designed
using the ``giant common block'' mentality, assuming that all data was
accessible directly from anywhere in the code (even if the implementation
didn't actually use common blocks).  This meant that the data structures
themselves were the interface, so that we couldn't restructure the data
without touching all of the code that uses it.
It also meant that we couldn't easily determine which data was used where,
for purposes of managing data movement between different memory spaces.
We hope that use of modern language features and better software
design techniques in our existing code bases will address these issues
and ease the transition to the newer architectures.

Similarly, when we considered code maintenance for the new codes we
were developing, we faced the prospect of having to maintain distinct
versions of large sections of the code for each new architecture we
were asked to support.  This would lead to nightmares in trying
to maintain consistency between an ever-increasing number of
implementations of similar algorithms.
Instead, we are trying to minimize this problem and reuse as much code as
possible on new architectures by improving the design of the software,
introducing appropriate abstractions, and using advanced language
features such as templating and function inlining.

Finally, we discovered several challenges related to software testing.
Most noticeably, we found that the increasing amount of concurrency
in the new architectures led to non-reproducible results on
a scale we had never experienced before.  For many years we had been
able to have consistent results as long as the hardware and software
environment remained constant; some variation could occur when
hardware, compilers, or library versions changed, but these changes
were infrequent and manageable.  On the new architectures, however,
variations could (and did) easily occur between one run and the next,
in ways that our usual testing methodologies weren't designed to handle.

To further complicate matters, we found that many of our legacy algorithms
were numerically unstable:  the smallest change to the inputs could
lead to large changes in outputs.  In the past developers did not pay much
attention to this issue, as it wasn't a problem; the small input changes
that exposed it were few and far between.  On new
architectures, however, the unstable algorithms would only amplify the
frequent small variations described above.
We found that we would need to add numerical stability
to our list of considerations in any new code we developed; and worse
yet, we would have to go back through thousands of lines of legacy
algorithms to determine their stability as well.

As a consequence of the non-reproducibility and non-stability we
discovered, our traditional testing methodology (create a reproducer,
rerun in the debugger, track down the error) would in
many cases be inadequate for future architectures.  We'd need to find other
ways of eliminating bugs from our code.  The good news was that modern
software practice provided a number of possible tools, such as:
\begin{itemize} \itemsep1pt \parskip0pt \parsep0pt
\item extensive unit testing, with particular attention to corner cases;
\item static and dynamic analysis tools; and
\item inspection techniques such as code reviews and pair programming.
\end{itemize}
The challenge, then, was to find ways to incorporate practices
like these into our current and future code projects.

Given all of these connections, it made sense for us to include a focus
on software practices as we continued our work on moving codes to
future architectures.

\section*{Two Experiments}

In the past few years,
LANL has run two small experimental projects as initial attempts
to raise awareness of new software practices and programming
approaches for new architectures.

First, starting in the fall of 2010, the ``Exascale Tutorial'' series
was launched.  This was a year-long, part-time ``boot camp'' for
experienced code developers from our large multi-physics code projects, 
intended to help them become more familiar with advanced architectures and
with modern software practices.  Four developers participated in
each of the fiscal years 2011 and 2012, with a few computer
scientists working alongside.
Work was done using C++ to program CPUs, and CUDA or OpenCL for GPUs;
most participants had little or no experience in these languages.
Attempts were made to practice requirements gathering, unit and
integrated testing, and documentation where appropriate.
The Exa11 group developed a small molecular
dynamics mini-app, first in a reference CPU version, then in several
GPU versions using various strategies and programming models.
The Exa12 group did similar work on other smaller codes and code fragments.
The participants generally described the tutorial as a useful educational
experience, and went on to use what they learned in other efforts;
many moved on to the SWIFT project described below.

Next, we launched the Software Infrastructure
for Future Technologies (SWIFT) project in the fall of 2011.  SWIFT
started with seven developers, representing a mix of domain science and
CS backgrounds, working half-time; this number increased
to ten in the second year.  The project goal was to use modern
software practices to produce, in four years, a prototype multi-physics
code suitable to run on future architectures.  (A much-repeated piece
of folklore at LANL states that ``it takes 15 years to develop a weapons
code''; thus, the four-year schedule for SWIFT was considered to be quite
ambitious.)

Modern software tools and techniques used on the SWIFT project included:
\begin{itemize} \itemsep1pt \parskip0pt \parsep0pt
\item C++ language for writing (nearly) all code;
\item Mercurial for distributed version control;
\item Eclipse for an integrated development environment;
\item cmake for a build system; and
\item CppUTest for a unit test harness.
\end{itemize}
No team member had used all of these before, and some had used
few or none of them.
In addition, the group used agile development practices adapted
from Alistair Cockburn's ``Crystal Clear''~\cite{CC}, including
co-location, pair programming, short (two-week) iterations, and daily
stand-up meetings.

The SWIFT project was terminated at the end of its second year, due
to tightening budgets and to pressing needs elsewhere in the program.
At that point, we had not made as much progress as we
had hoped toward the four-year goal (some reasons will be explored below).
However, we had developed some design ideas that were demonstrated in
several prototype codes, and learned useful lessons in a number of areas
along the way.
The SWIFT team members are now being deployed to other code projects
to help with modernization efforts, in hopes that the lessons learned
in SWIFT can be applied to those projects.

\section*{Lessons Learned}

Participants in both of the experimental projects gained valuable
experience with new techniques and tools.  We found some of these to be
helpful and applicable to LANL projects; others were perhaps helpful
in principle, but had enough problems and limitations that we would
recommend alternatives for future projects.
We also had our first experiences at LANL with the disciplined
development practices that are common in the software industry.
Since these were new to us, we understandably encountered many
rough spots, and our execution could have been better.  Still, we gained
enough positive experience to agree we should continue in this
direction, learning more about these practices and finding ways
to use them at LANL.

In addition to the technical knowledge we gained, we
benefitted from the increased interaction between domain scientists and
computer scientists.  Historically, these two communities at LANL have
for the most part operated in isolation.  The projects
gave us good opportunities to learn from each other, and to build
working relationships that are continuing into other efforts.
It also helped team members from each of the groups (domain science and CS)
to realize that the other group had something important to contribute
to the scientific code development process.

There were issues, though, that kept these projects from being as
productive as they perhaps could have been.
One disadvantage, noted by several participants in
both projects, is that it's difficult for a learner to stretch in several
different directions simultaneously.  The domain scientists in
particular were often learning a new language, new tools, new kinds
of architectures, and in some cases even a new problem domain, all at
the same time; this was sometimes too much to take in at once.

Another disadvantage to both projects was their dual nature:
were we supposed to be experimenting and free to fail, so we could learn
from our failures?  Or were we working toward a normal deliverable with
a hard deadline?
Often we tried to do both, and as a result we didn't do either
very well.  This caused some difficulty in particular with our use
of agile development methods, which are designed for ``real'' projects;
we couldn't always see how to make these fit in ``experimental'' mode.
Even so, we generally agreed that the basic agile
ideas were better suited to the LANL development environment than the
more formal, structured project lifecycles that some team members had
used elsewhere.

We also discovered some larger cultural issues at LANL that made it
hard to fully implement our agile development ideas.  Since all of the
experimental projects were only part-time efforts, we attempted to
co-locate for whatever fraction of time we were working, to allow
for things like pair programming, group design reviews, and extended
brainstorming sessions.  We found this to be a useful practice.
However, the LANL culture tends to be very much interrupt-driven:
it's expected that people, especially key people, are available on
demand.  There was a common perception that our experimental
work didn't have ``real'' deadlines, so that other deliverables and
milestones took precedence.
This made it hard to protect our co-location time against interruptions,
and our productivity suffered as a result.

\section*{Next Steps}

Efforts are underway in several of our large code projects to
modernize the code base and prepare for future architectures.
As noted earlier, the SWIFT project members will be deployed to
help in these efforts, bringing with them the lessons they have
learned related to sustainable software practices.
Some higher-level discussions with management will be needed
to try and address larger cultural issues that were
exposed in the tutorials and the SWIFT project.

In the past couple of years we have also started sharing our
experiences with the other NNSA labs (Sandia and Livermore)
as they have ramped up on advanced architecture work.  As more
labs and universities begin to deal with these architectures,
we hope to expand our level of collaboration with them
in this area.


\begin{thebibliography}{9}
\bibitem{CC}
Cockburn, Alistair. {\em Crystal Clear:  A Human-Powered Methodology
for Small Teams}, Addison-Wesley, Boston, 2005.
\end{thebibliography}
\end{document}